\documentclass[]{spie}  

\usepackage{amsmath,amsfonts,amssymb}
\usepackage{graphicx}

\title{In-orbit Performance of UVIT on ASTROSAT}

\author[a]{Annapurni Subramaniam}
\author[a]{Shyam N. Tandon}
\author[b]{John Hutchings}
\author[c]{Swarna K. Ghosh}
\author[a]{Koshy George}
\author[d]{V. Girish}
\author[a]{P.U. Kamath}
\author[a]{S. Kathiravan}
\author[a]{Amit Kumar}
\author[a]{J. Paul Lancelot}
\author[a]{P.K. Mahesh}
\author[a]{Rekhesh Mohan}
\author[a]{Jayant Murthy}
\author[a]{S. Nagabhushana}
\author[a]{Ashok K. Pati}
\author[e]{Joe Postma}
\author[a]{N. Kameswara Rao}
\author[d]{Kasiviswanathan Sankarasubramanian}
\author[a]{P. Sreekumar}
\author[a]{S. Sriram}
\author[a]{Chelliah S. Stalin}
\author[a]{Firoza Sutaria}
\author[a]{Yuvraj Harsha Sreedhar}
\author[a]{Indrajit V. Barve}
\author[a]{Chayan Mondal}
\author[a]{Snehalata Sahu}

\affil[a]{Indian Institute of Astrophysics, Koramangala II Block, Bangalore-560034, India}
\affil[b]{National Research Council, Canada}
\affil[c]{National Center for Radio Astrophysics, Pune, India}
\affil[d]{ISRO Satellite Centre, HAL Airport Road, Bangalore 560017}
\affil[e]{University of Calgary, Canada}

\authorinfo{Further author information: (Send correspondence to Annapurni Subramaniam)\\E-mail: purni@iiap.res.in }

\begin{document} 
\maketitle

\begin{abstract}
We present the in-orbit performance and the first results from the ultra-violet Imaging telescope (UVIT) 
on ASTROSAT. UVIT consists of two identical 38cm coaligned telescopes,  one for the FUV channel (130-180nm) and the 
other for the NUV (200-300nm) and VIS (320-550nm) channels, with a field of view of 28 $arcmin$. The FUV and the NUV
detectors are operated in the high gain  photon counting 
mode whereas the VIS detector is operated in the low gain integration mode. The FUV and NUV channels have filters and gratings,
whereas the VIS channel has filters. The ASTROSAT was launched on 28th September 2015.  
The performance verification of UVIT was carried out after the opening of the UVIT doors on 
30th November 2015, till the end of March 2016 within the allotted time of 50 days for calibration.
 All the on-board systems were found to be 
working satisfactorily. During the PV phase,
the UVIT observed several calibration sources to characterise the instrument and a few objects to 
demonstrate the capability of the UVIT. The resolution of the UVIT was found to be about 1.4 - 1.7 $arcsec$ in the FUV and NUV.
The sensitivity in various 
filters were calibrated using standard stars (white dwarfs), to estimate the zero-point magnitudes as well as the flux conversion factor.
The gratings were also calibrated to estimate their resolution as well as effective area.
The sensitivity of the filters were found to be reduced up to  15\% with 
respect to the ground calibrations. The sensitivity variation is monitored on a monthly basis. At the end of the PV phase,
the instrument calibration is almost complete and the remaining calibrations will be completed by
September 2016. UVIT is all set to roll out science results with its imaging capability with good resolution and large field of view,
 capability to sample the UV spectral region using different filters and capability to perform variability
studies in the UV. 

\end{abstract}

\section{Introduction}
\label{sec:intro}  

Ultra Violet Imaging Telescope (UVIT) is one of the 5 payloads on board the first Indian space observatory, ASTROSAT. 
ASTROSAT is a multi-wavelength mission which can observe simultaneously from hard X-ray to optical wavelength. The other
4 payloads are the Large Area X-ray Proportional Counters (LAXPC), Cadmium-Zinc-Telluride Imager (CZTI), Soft X-ray Imaging Telescope (SXT)
and Scanning Sky Monitor (SSM). The primary science objectives of ASTROSAT are 
(1) Multi-wavelength study of astronomical sources
(2) High energy processes in accreting compact objects and extragalactic systems
(3) Measuring magnetic fields of neutron stars
(4) Search for transient sources
(5) Stars and star formation in nearby galaxies 
\\

\section{UVIT - Telescope}
\label{sec:tel}  

The UVIT instrument contains two 38 cm telescopes: one for the far-ultraviolet (130 - 180 nm : FUV); and the second for the near-ultraviolet (200-300 nm  : NUV) and the visible (320 - 550 nm  : VIS) ranges, which 
uses a dichroic mirror for beam-splitting. UVIT is primarily an imaging instrument, which simultaneously makes images in the FUV, NUV and VIS
channels, with 
 a field of view of  ~ 28 $arcmin$. 
Each of these channels are divided into smaller pass bands, which can be selected using a 
set of filters. In addition, a slit-less spectroscopic mode is available in FUV and NUV. 

The primary mirror of theFUV telescope has a working diameter of 375 mm.
The detector has a diameter of $\sim$ 40 mm. The filters, each 
of diameter 50 mm, and the grating are mounted in a wheel 
at a distance of  $\sim$ 40 mm  from the detector.
Optical layout of the NUV /VIS telescope is similar to the FUV telescope. 
A dichroic beam splitter is used for spectral division of the beam 
in NUV (reflection) and VIS (transmission). 
Photon counting detectors are used for FUV and NUV channels. 
In order to correct for drift of the
satellite and achieve the required spatial 
resolution, short exposures are taken in the VIS channel and are integrated through a shift and 
add algorithm on the ground. 
The drift between the consecutive frames in VIS channel is computed and the resulting relative aspect time series 
data is used to correct for the drift in NUV and FUV channel data.
More details regarding the telescope can be found in the ASTROSAT hand book 
(http://astrosat.iucaa.in/).

\section{Detector and the filter system}
\label{sec: det}  
The UVIT Detectors are of photon counting nature based on Micro Channel Plate
(MCP) image intensifiers Technology. The photocathode deposited on the inside of a 
40 mm window lies at the focal plane of the optical system. The photoelectrons 
released are accelerated across a gap (typically 100-200 $\mu$m) to a stack of two
micro-channel plates. The resulting electron shower illuminates a
phosphor with a fast decay time. The light from the phosphor is fed
through a fiber-optic taper that is bonded to the surface of a Cypress
Star250 CMOS detector, with 512 $\times$ 512 pixels. Each pixel is square in
size with 25 $\mu$m a side. Each pixel extends $\sim$ 3$\times$3 $arcsec$ on the
sky. A detected photon event produces a splash of light on the CMOS
that covers several pixels. The exact coordinates of the photon event
would be estimated through a centroid algorithm using the
pixel values of the detector, to a much higher resolution than one CMOS
pixel. The experimental studies done by Hutchings et al. (2007) show
that one photon event produces a light splash which follows roughly a
Gaussian distribution with Full Width at Half Maximum (FWHM) of $\sim$
1.5 pixels.  The UVIT detectors
are designed with the gap of 0.1 – 0.15 mm, between the photocathode
and the MCP, to obtain a resolution of $\sim$
1" ($\sim$ 23 $\mu$m on the photo-cathode).
The intensified CMOS detectors can either be used in a high gain photon counting mode, 
or in a low gain integration mode (in which signal in a pixel of the CMOS detector 
could be contributed by multiple photons). The detector can also be used in window mode
for observing partial fields to get faster frame rates ($>$ 29/s). For a window size
of 100 $\times$ 100 pixels (5.5 $\times$ 5.5 arcmin$^2$), a frame rate of 600 frames/sec is possible. 

The UVIT has various filters in the three channels. 
Each filter wheel also holds a block which will be used to avoid bright objects and
to protect the detector.  The gratings have a resolution of about 100 and operate in a mode similar to 
slit-less spectroscopy.

\section{Ground calibrations}
The calibration of the UVIT instrument is achieved through ground-based as well as in-orbit calibration tests/observations.
The individual components of the UVIT instrument are tested and calibrated in the ground. The performance of the integrated system is 
verified during the performance verification phase after the launch, in orbit.  

\subsection{Ground calibration tests}
1. Detectors:\\
The CMOS detectors used in the three UVIT channels are tested at the Canadian facility in the University of Calgary and MGK Menon Lab, IIA.  
The tests performed on the detector include tests for sensitivity, response as a function of wavelength, spatial variation of sensitivity, tests to estimate the centroid of the photons detected, gain as a function of MCP voltage etc. Some results of
the above tests are summarised in Postma, Hutchings \& Leahy (2011). The quantum efficiency and distortion of the detectors
were estimated at the MGK Menon Lab of IIA.
Details of the set up, analysis and the results of the estimation
are presented in the Technical report (Narra et al. 2011a). 
The EM and the FM filters and gratings were tested and calibrated at the 
MGK Menon Lab. The parameters 
estimated were the transmission of each filter as a function of wavelength, spatial 
variation of sensitivity, parallelism of the sides, shift in focus due to the filters, etc.
Details of the experiment, analysis and results for the filters 
are presented in technical report (Shankarasubramanian et al. 2011). The report containing the 
results on grating can be found in Narra et al. 2011b. \\
The following parameters of the detectors, filters and grating were calibrated in the ground. Main results of the ground calibration tests are available in the UVIT-CDR documents (http://www.iiap.res.in/Uvit).\\
\noindent 1. Effective wavelength and bandwidth of the filters\\
2. Response and sensitivity\\
3. Spatial variation of sensitivity as a function of wavelength\\
4. Photometric non-linearity in Photon counting  mode\\
5. Effects of window mode on sensitivity\\
6. Gain vs MCP voltage in integration mode\\
7. Distortions on the detector\\
8. Efficiency, wavelength coverage of gratings\\
9. Length of spectra on the detector plane\\
10. Shift between the object position in the image and the zeroeth order spectrum\\

Figures 1, 2 and 3 show the effective area curve of the telescope for the filters in the three channels, FUV, NUV and VIS, respectively, 
derived based on ground calibrations. \\

   \begin{figure} [ht]
   \begin{center}
   \begin{tabular}{c} 
   \includegraphics[height=7cm]{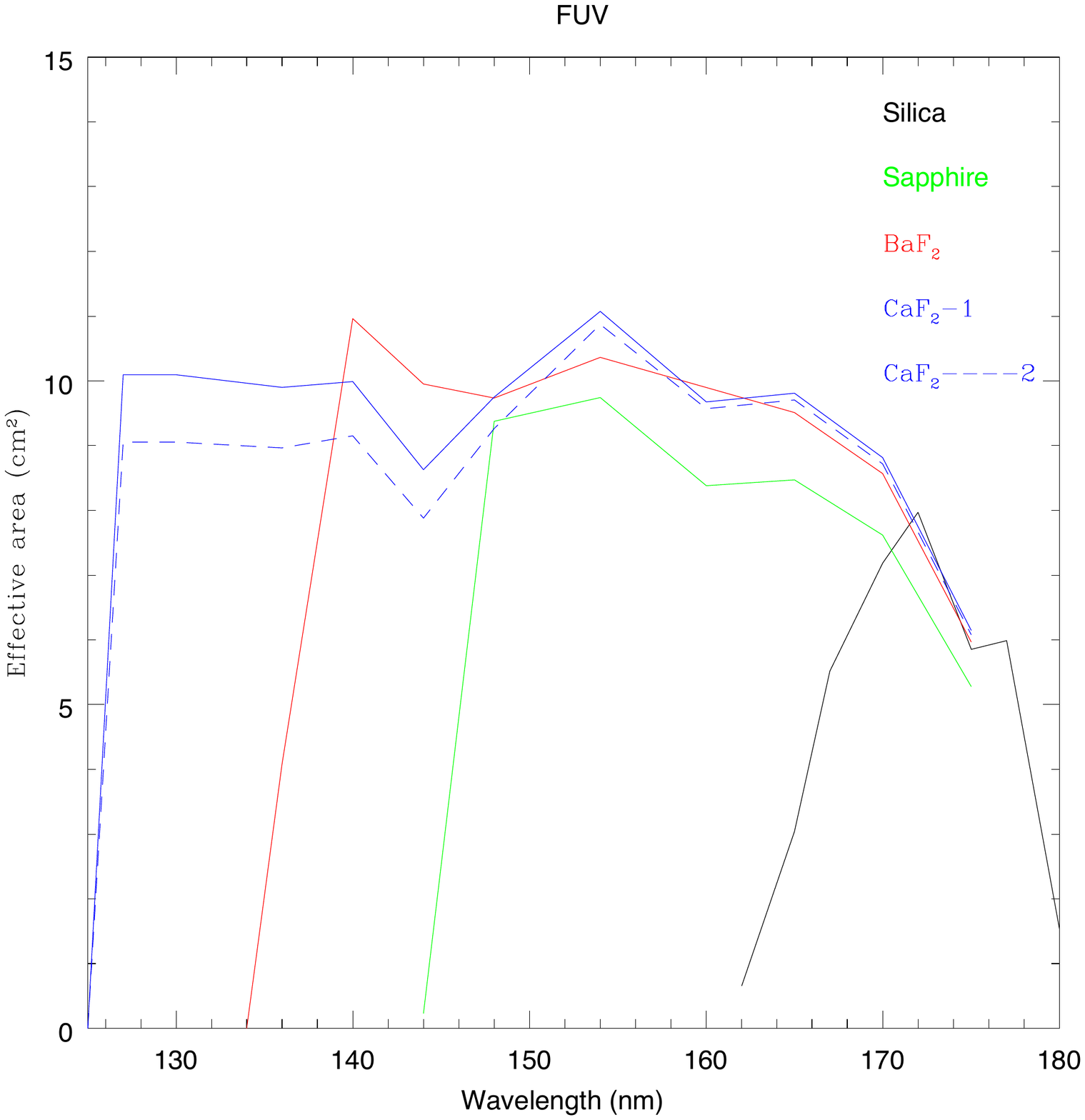}
   \end{tabular}
   \end{center}
   \caption[example] 
   { \label{fig:eff_fuv} 
Effective area of the telescope for the FUV filters of UVIT. These are derived from the ground calibrations.}
   \end{figure} 

   \begin{figure} [ht]
   \begin{center}
   \begin{tabular}{c} 
   \includegraphics[height=7cm]{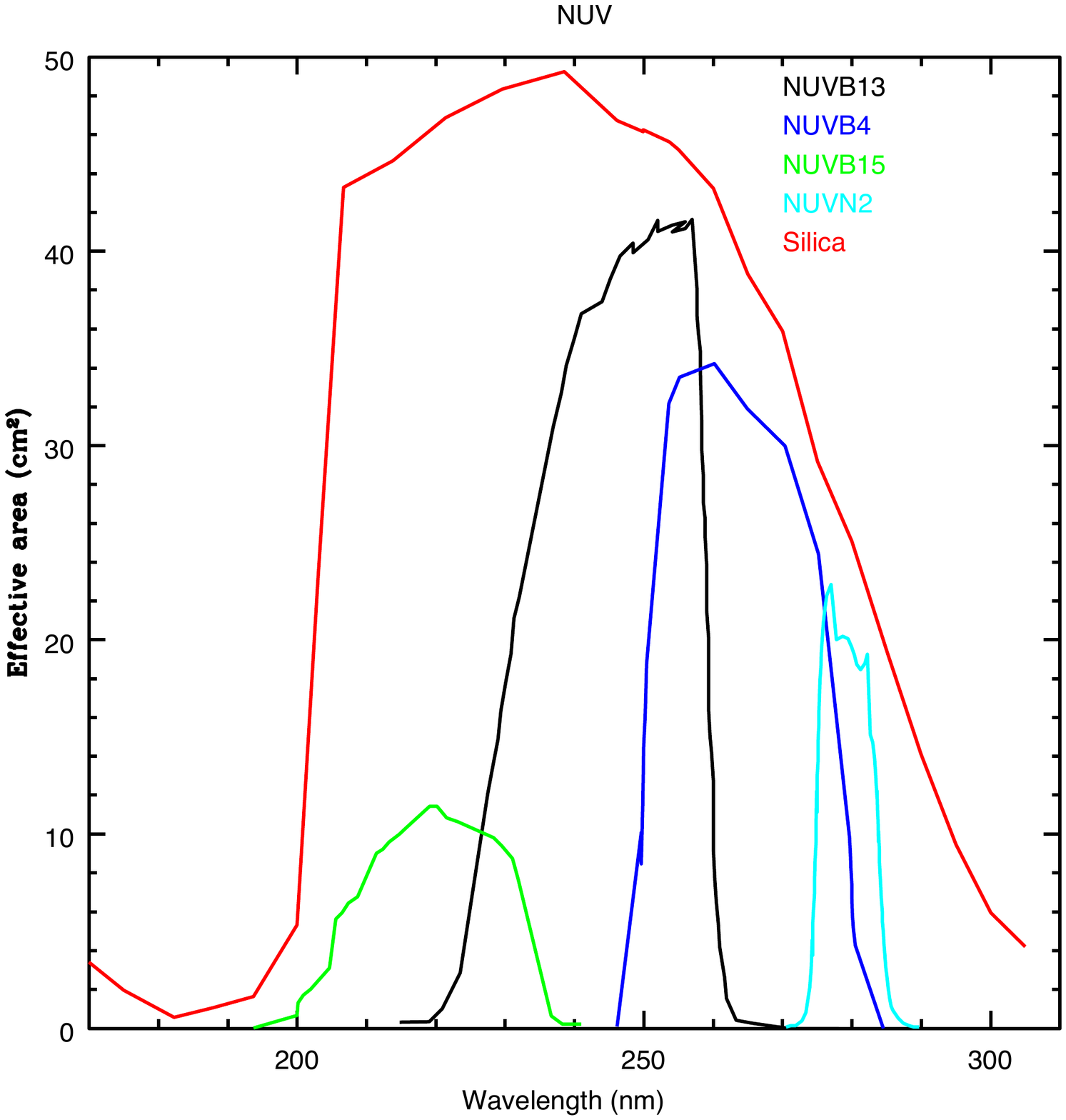}
   \end{tabular}
   \end{center}
   \caption[example] 
   { \label{fig:eff_nuv} 
Effective area of the telescope for the NUV filters of UVIT. These are derived from the ground calibrations.}
   \end{figure} 

   \begin{figure} [ht]
   \begin{center}
   \begin{tabular}{c} 
   \includegraphics[height=7cm]{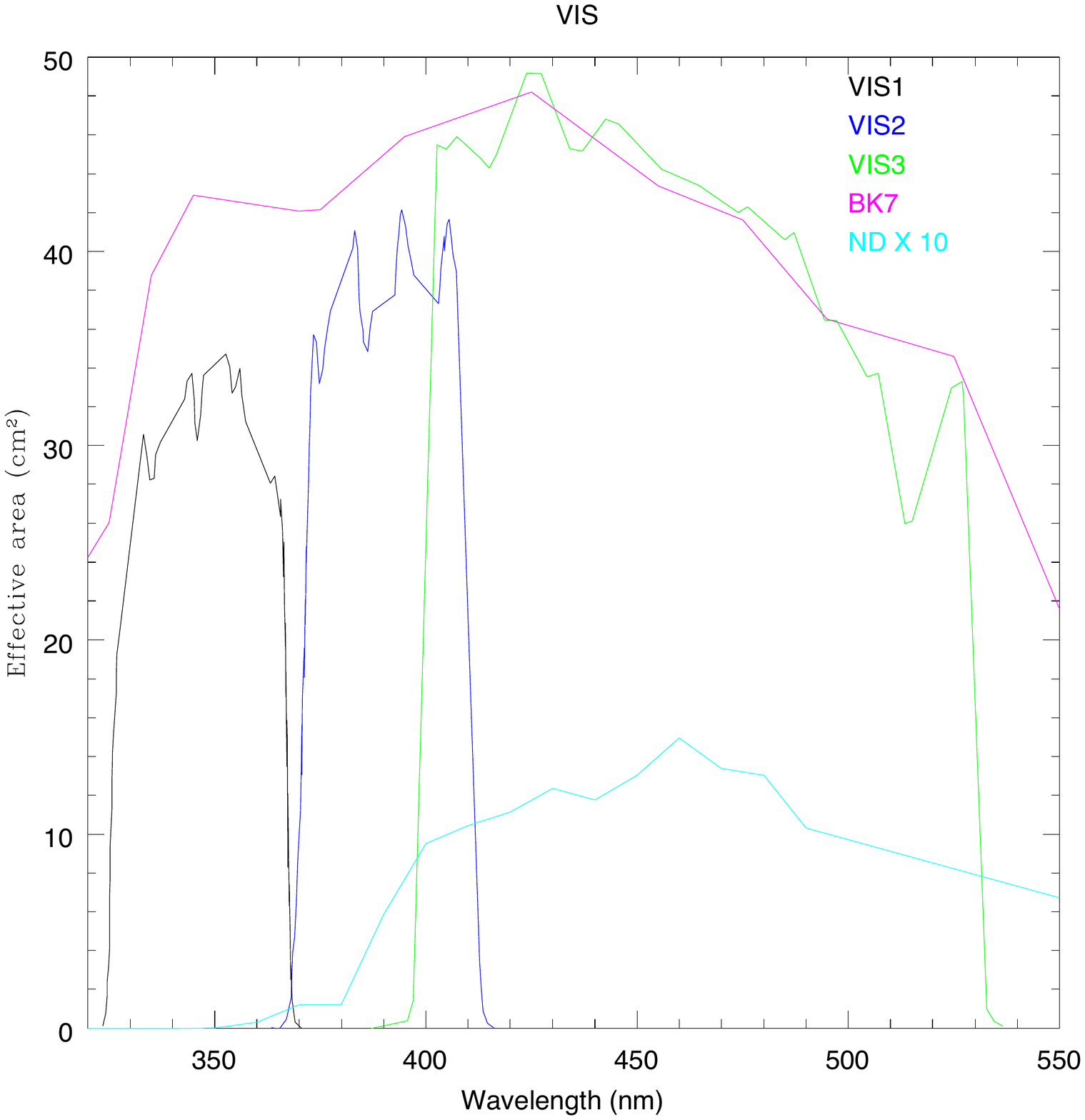}
   \end{tabular}
   \end{center}
   \caption[example] 
   { \label{fig:eff_vis} 
Effective area of the telescope for the VIS filters of UVIT. These are derived from the ground calibrations.}
   \end{figure} 

\section{In-orbit calibrations}
A calibration task is the configuration, observation and analysis required to calibrate a specific aspect of the instrument. 
The primary goal of this calibration is to ensure that the scientific quantities obtained from the UVIT instrument conforms 
to the requirements set forth by the instrument document. This also needs to fulfil the requirements set forward by the science projects. 

ASTROSAT
was launched on 28th September 2015, from Satish Dhawan Space Centre (SDSC), India by the 
Indian Space Research Organisation (ISRO). It is placed in an equatorial orbit
with 6$^o$ inclination, at an altitude of 650km.
 After the launch, the
payloads were switched on one after the other, with UVIT coming on at the last. The electronics of the UVIT was switched on after 6 days,
where the power was switched on one channel at a time. The electronic tests were completed by 13th November 2015. The doors of the UVIT
were opened on 30 November 2015, this operation was performed 62 days after the launch of ASTROSAT. The first target for observation was
chosen as NGC 188, an old open cluster located at high declination. The initial observations were carried out on this object, before
moving on to other targets. Analysis of the images indicated that a good resolution was obtainable after corrections for the drift 
and that the sensitivity in the far ultraviolet (130-180 nm) was good indicating that no major contamination had occurred during the 
launch or in the orbit.  As this object, NGC 188, is located in the clear viewing zone, this target will be available throughout the year and
is used to check the sensitivity of UVIT, with observations taken every month.

The performance verification (PV) phase of the UVIT is of approximately 50 days duration during the
window of 6 months after the launch, till March 2016.
Various calibration tasks planned for the PV phase and their completion status are listed in Table 1. Some tasks are completed whereas the
others are in progress, which mostly means that the analysis of the observations are in progress. The task which needs observation
is the calibration of the Silica filter in the NUV channel. This will be carried out in the coming months.
\begin{table}[ht]
\caption{The planned and completed PV phase calibrations are listed here. }
\label{tab:fonts}
\begin{center}       
\begin{tabular}{|l|l|} 
\hline
\rule[-1ex]{0pt}{3.5ex}  Photometric Zero-point estimation (FUV filters) & All completed \\

\hline
\rule[-1ex]{0pt}{3.5ex}  Photometric Zero-point estimation (NUV filters) & 4 completed, one pending \\
\hline
\rule[-1ex]{0pt}{3.5ex}  Flux calibration of Grating (FUV and NUV) & completed   \\
\hline
\rule[-1ex]{0pt}{3.5ex}  Dispersion and wavelength calibration of Grating (FUV and NUV) & completed   \\
\hline
\rule[-1ex]{0pt}{3.5ex}  Calibration of detector distortion & in progress   \\
\hline
\rule[-1ex]{0pt}{3.5ex}  Calibration of saturation & in progress  \\
\hline
\rule[-1ex]{0pt}{3.5ex}  PSF estimations & complete  \\
\hline
\rule[-1ex]{0pt}{3.5ex}   Timing calibration & in progress \\
\hline
\rule[-1ex]{0pt}{3.5ex}  Estimation of background & completed  \\
\hline
\rule[-1ex]{0pt}{3.5ex} Secondary photometric calibration & regular monitoring in progress \\
\hline
\end{tabular}
\end{center}
\end{table} 

\subsection{Photometric calibrations}
The Photometric calibration of the UVIT filters is one of the important activity in the in-orbit calibration phase. 
The primary photometric calibration of the UVIT filters can be achieved by observing standard stars for which flux 
calibrated spectra are available in the UV wavelength range. The deliverables of the primary photometric calibration 
are the Zero Point magnitude for all the filters and the unit conversion factor. The magnitude system adopted for the UVIT filters 
is the AB magnitude system (Oke 1990) and hence the magnitudes derived will be in this system. The unit conversion
 is the flux conversion factor to derive the flux of the source in each filter at a particular wavelength. 
The primary photometric standard star should have flux calibrated spectrum available in the wavelength range covering 
the FUV and NUV filters of the UVIT. Such sources are available in the CALSPEC database of the HST and these are the 
potential targets for the primary photometric calibration. As these sources have flux calibrated spectra, we can 
predict the expected count rates based on the ground calibrations. 
From the ground calibrations, we have the effective area, mean and effective wavelengths and band pass of the filters.  
The expected Zero Point magnitude and the unit conversion values were estimated based on the ground calibrations. 
A comparison of the estimated count rates of standard star with the observed count rates will estimate the efficiency
of the telescope in the filter system. 

The equations governing conversions between Zero Point magnitude, Unit Response and Count per second (CPS) are given below. 
For each filter, the Flux as well as the AB magnitude are defined at the corresponding $\lambda$ (mean)

\begin{equation}
UVIT Flux (ergs/sec/sq.cm/{\AA}) = Unit Response *CPS 
\end{equation}

\begin{equation}
UVIT magnitude (AB) = -2.5log10(CPS) + Zero Point Magnitude
\end{equation}

The selection of the standard star for the PV phase, from the planning document is based on a few factors. 
Most importantly, the source should be available during the PV phase. The count rate of the star should be 
less than 30 counts per second (to reduce effects due to saturation) in any filter and it should also
 be not less then a few counts per second (to get good signal). 
The field of view should not have any bright star which could trigger Bright object detection (BOD), which in turn triggers a
shut down of the system. Based on the above constraints, two standard stars were shortlisted for observations 
in the PV phase. These two standard stars are HZ4 and LB227. 

HZ4 is suitable for observations in all the FUV filters, whereas it has relatively high count rates in the NUV broad band filters. 
Therefore, it was observed in only two filters (NUVN2 and NUVB15) in the NUV channels. As there are some bright 
objects in the field, the VIS channel was configured for ND (Neutral density) filter. 
The spectrum of HZ4 was convolved with the effective area for the filters and the expected count rates and 
magnitudes were estimated. 

The source HZ4 was observed on 1st and 2nd February 2016. The field was observed in the above mentioned filters. The data were
obtained at the rate of 29 frames per sec and the effective exposure times were found to range between 15 min to 5 min.
Each frame was corrected for drift and were merged together to create images. The frames were also corrected for flat, using flat
fields derived from the ground calibration. 
The images, thus created in the observed filters, were then used to estimate count rates using aperture photometry, 
after correcting for the background.
These estimated count rates need to be corrected for saturation. To estimate the saturation, it is assumed that multiple 
photon events for the source (in a frame) are detected as single photon event or the corrected photon rate per frame 
is “X” where $(1- exp(-X))$ is the observed rate. The saturation corrected count rates are then compared with the expected count rate from HZ4.

The difference between the observed CPS (after correcting for saturation) and the expected CPS, is due to the
change in the effective area. This change in the effective area is likely due to the reduction in the
sensitivity of the components of the telescope. This is found to be up to 15\% for various filters, except for the filter, NUVB15, which
is found to show a reduction of about 50\% in its sensitivity. This factor is then used to estimate the
revised effective area of the system, which is tabulated in table 2. The zero point magnitude and the unit conversion are then estimated again
using the new effective area, and these are also tabulated in the table 2. We notice a relatively large reduction in the sensitivity of CaF2 and
BaF2 filters, when compared to the other filters in the FUV. This may be due to the contamination in the shorter wavelength 
region of these filters, which are more sensitive to contamination.

The observations in the other standard star LB 227 were also carried out. As there are a few bright sources in the field of this star,
we decided to offset this star from the centre, for safe observations. The observations were carried out in all the filters except, 
the broad band Silica filter in the NUV. The estimated count rates in all the filters (corrected for saturation effects), suggest similar
reduction in sensitivity. The two NUV filters, NUVB13 and NUVB4 (other than NUVN2 and NUVB15) were used for these observations and the reduction in the sensitivity were also
estimated. Though the zero point magnitude and unit response are estimated, we have not listed the values in the table.
As these images have the standard star off-centered, the analysis would require better flat fielding correction. 
We expect to refine the estimated values soon, based on more accurate flat field images.

\subsubsection{Spectroscopic calibrations}
The UVIT has two gratings in the FUV channel and one grating in the NUV channel. The two gratings in the FUV channel are placed orthogonal 
such that the dispersion axes are perpendicular. These gratings were calibrated using standard stars in the in-orbit phase. The calibrations
include the estimation of wavelength range, dispersion relation and location of various orders. The object observed for this purpose, was the
bright planetary nebula, NGC 40. These observations were obtained on 10th December 2015. The dispersion estimated for the NUV grating
is found to be 44.5\AA / pixel (for the first order) and 22.6 \AA / pixel (second order) for the FUV grating.  These correspond to a 
resolution of $\sim$ 90 and $\sim$ 110 for the NUV and FUV respectively.

The flux calibration of the gratings were performed using the standard star HZ4. The observations were obtained on 3rd February 2016 in the
NUV and FUV gratings. These observations were used to estimate the effective area of the gratings. The estimated curves are shown in the figure 4 and 5 
for the NUV and the FUV gratings respectively.

\subsubsection{PSF}
The point spread function of the NUV and FUV channels were estimated using drift corrected images with good signal to noise ratio. 
The PSF in the FUV and NUV channels are found to be in the range, 1.4 -- 1.7 $arcsec$. This is found
to be better than the specification, which is $\sim$ 1.8 $arcsec$, which is three times better than the resolution of GALEX.

\subsubsection{Background}
In the FUV band, the Geocoronal lines of OI at 130.4nm and 135.6nm are expected to give about 700 cps and 
50cps for the CaF2 and BaF2 filters respectively. This is the main source of background in the FUV, apart
from the direction dependent background from the ISM. The observed background for CaF2 -1 is about 200 cps,
in the field of NGC 2336, of which about 120cps could be due to contribution from cosmic rays. 
The NUV background is primarily due to Zodiacal light and us about 25 mag/10 sq.arcsec, with a variation
of up to 2 mag with helio-ecliptic latitude.

\begin{table}[ht]
\caption{The calibration results for various filters are listed below. The parameters listed are the mean wavelength of the 
filter and the pass bands (in \AA), the effective area (sq.cm), corrected after the in-orbit calibrations, 
the unit response and the zero point magnitudes. For comparison, the corresponding values of the GALEX mission are also listed.} 
\label{tab:Paper Margins}
\begin{center}       
\begin{tabular}{|l|l|l|l|l|l|} 
\hline
\rule[-1ex]{0pt}{3.5ex}  Filter & $\lambda$ (mean) & $\delta \lambda$ & EEA & Unit Response & Zero Point  \\
\hline
\rule[-1ex]{0pt}{3.5ex} {\bf UVIT - FUV} & & & & &  \\
\hline
\rule[-1ex]{0pt}{3.5ex}  CaF2-1 & 1480.8 & 500 & 9.21 & 0.29153E-14 & 18.08  \\
\hline
\rule[-1ex]{0pt}{3.5ex}  BaF2 & 1540.8 & 380 & 9.83 & 0.26921E-14 & 17.80  \\
\hline
\rule[-1ex]{0pt}{3.5ex}  Sapphire & 1607.7 & 290 & 10.08 & 0.42305E-14 & 17.50  \\
\hline
\rule[-1ex]{0pt}{3.5ex}  Silica & 1716.5 & 125 & 8.96 & 0.18696E-13 & 16.38  \\
\hline
\rule[-1ex]{0pt}{3.5ex} {\bf UVIT - NUV} & & & & &  \\
\hline
\rule[-1ex]{0pt}{3.5ex}  NUVB15 & 2195.5 & 270 & 12.31 & 0.54151E-14 & 16.55  \\
\hline
\rule[-1ex]{0pt}{3.5ex}  NUVN2 & 2792.3 & 90 & 24.55 & 0.36443E-14 & 16.46  \\
\hline
\rule[-1ex]{0pt}{3.5ex} {\bf GALEX} & & & & &  \\
\hline
\rule[-1ex]{0pt}{3.5ex}  FUV & 1538.6 &  & 19.6 & 0.140E-14 & 18.82  \\
\hline
\rule[-1ex]{0pt}{3.5ex}  NUVN & 2315.7 &  & 33.6 & 0.206E-15 & 20.08  \\
\hline
\end{tabular}
\end{center}
\end{table}

\begin{figure} [ht]
\begin{center}
\begin{tabular}{c} 
\includegraphics[height=7cm]{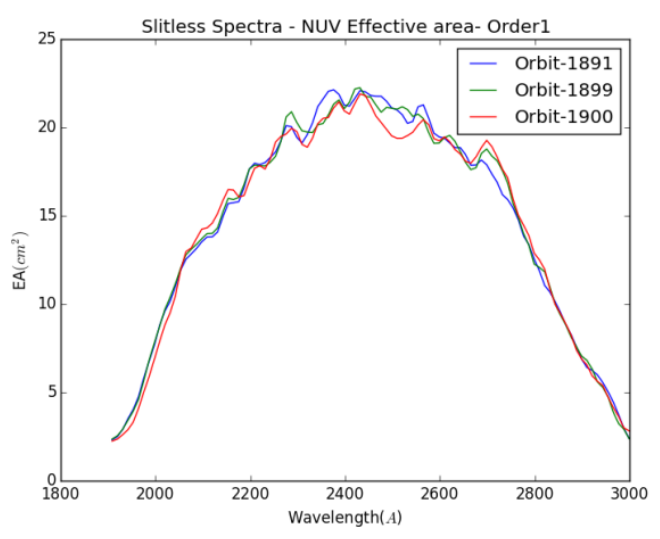}
\end{tabular}
\end{center}
\caption[example] 
   { \label{fig:example} 
The effective area of the NUV grating in the first spectral order.}
   \end{figure} 

\begin{figure} [ht]
\begin{center}
\begin{tabular}{c} 
\includegraphics[height=7cm]{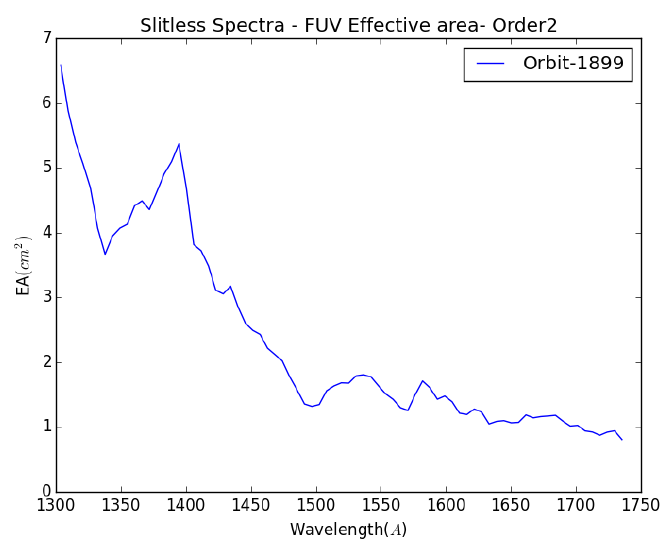}
\end{tabular}
\end{center}
\caption[example] 
   { \label{fig:example} 
The effective area of the FUV grating in the second spectral order.}
   \end{figure} 

\section{Observations to demonstrate the capability of UVIT}
Along with the observations to perform  calibrations, a few objects were observed to demonstrate the capability of UVIT, during the
PV phase. A summary is given below.

\begin{itemize}

\item[1.] The observations of a star forming spiral galaxy, NGC 2336 was carried out to demonstrate the capability of
UVIT to image with good resolution. These observations were carried out in the Silica filter in the NUV and CaF2 filter in the FUV.
A sample image in the NUV for an integration time of about 25min is shown in the figure, along with the image taken by GALEX, for comparison.

\begin{figure} [ht]
\begin{center}
\begin{tabular}{c} 
\includegraphics[height=9cm]{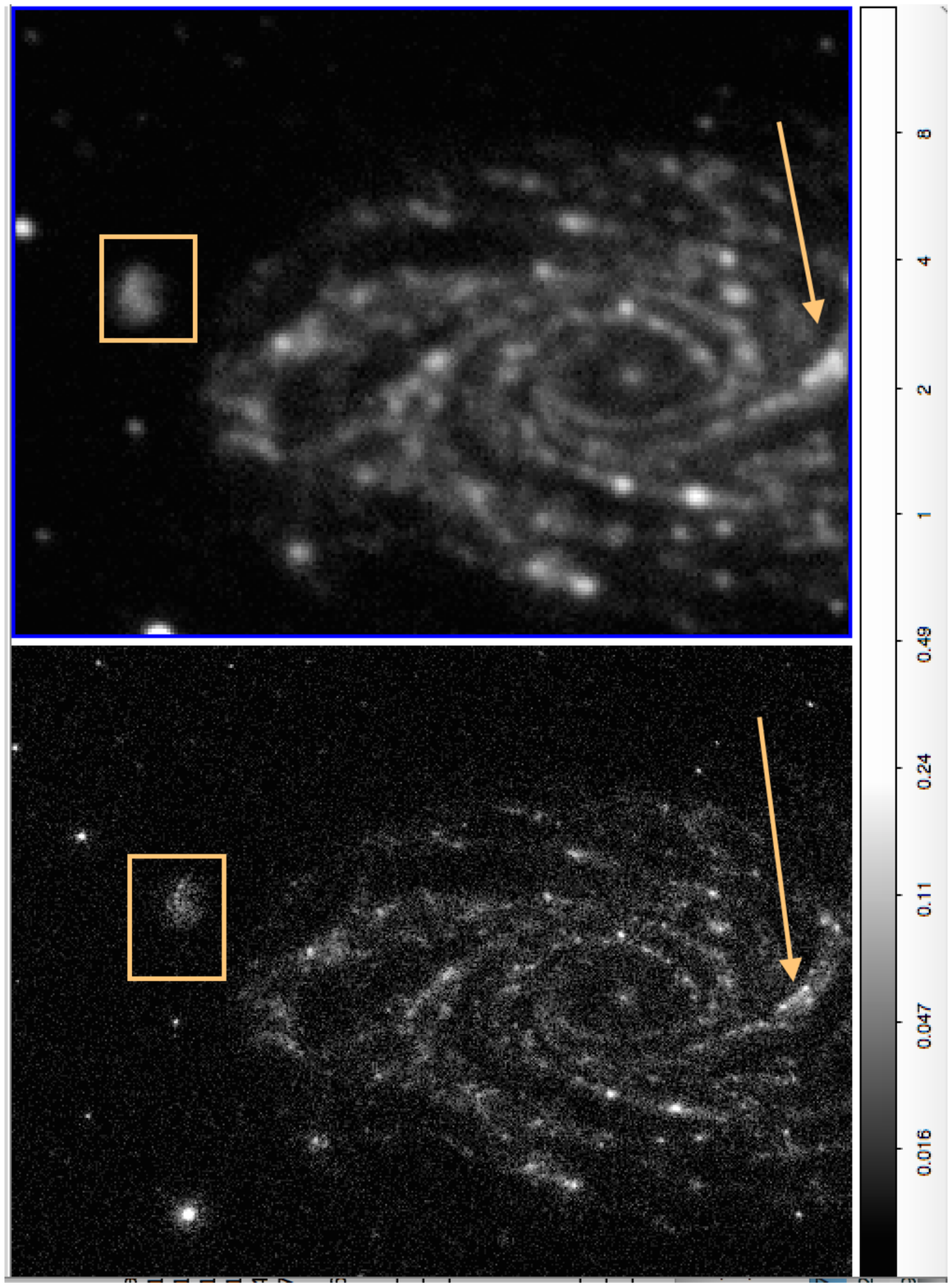}
\end{tabular}
\end{center}
\caption[example] 
   { \label{fig:example} 
The image of NGC 2336 obtained by UVIT in the NUV channel using Silica filter is shown below. For comparison, the image obtained
by GALEX is shown above. Specific regions of the galaxy, where UVIT image has resolved the star forming knots are indicated.}
   \end{figure} 

\item[2.]The rich stellar field of the open cluster NGC 188 taken using the NUVN2 filter in the NUV channel is also shown, to demonstrate
the imaging capability of UVIT. The cluster is found to have a good number of stellar sources detected in the NUV region. 

\begin{figure} [ht]
\begin{center}
\begin{tabular}{c} 
\includegraphics[height=7cm]{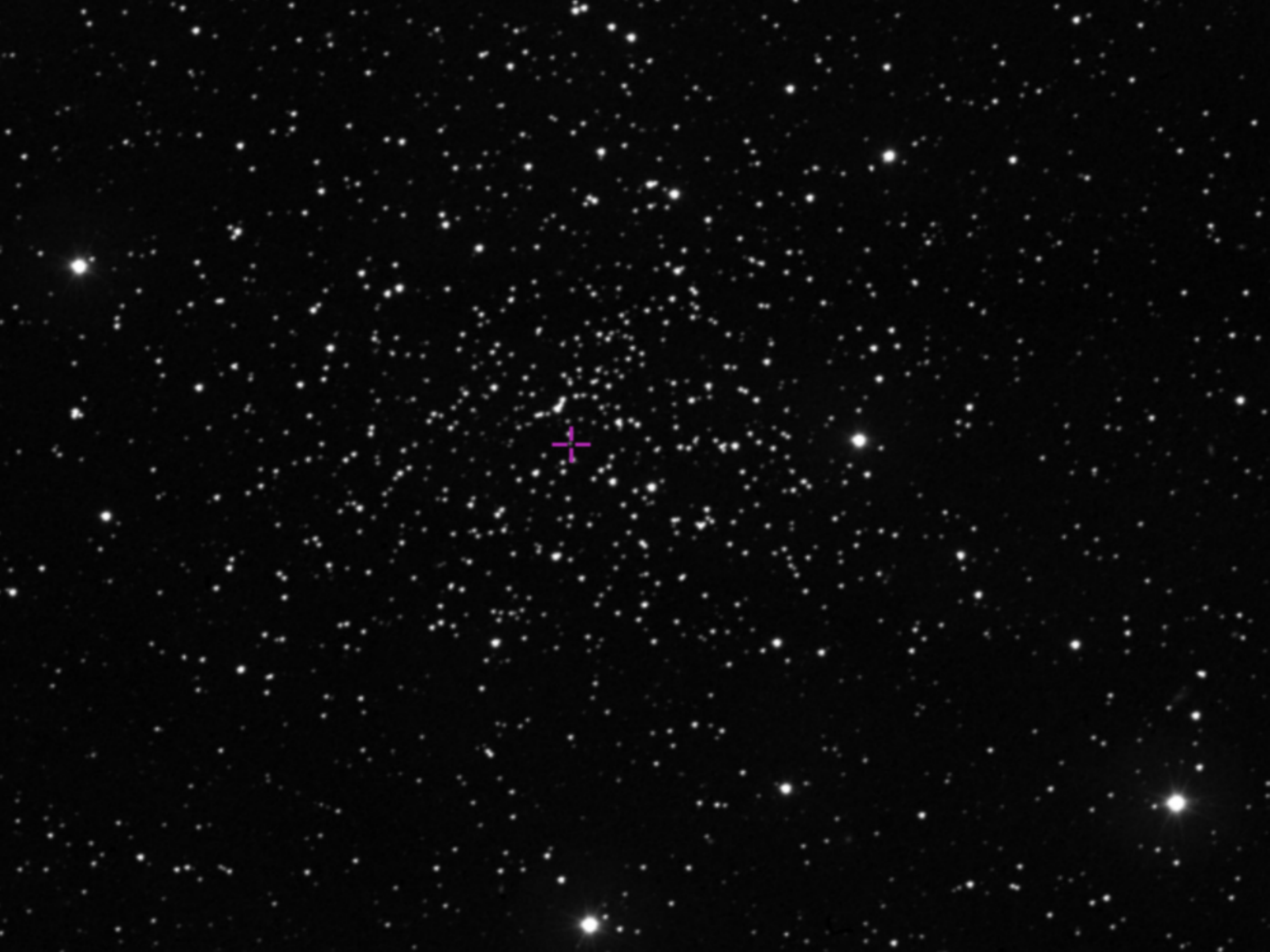}
\end{tabular}
\end{center}
\caption[example] 
   { \label{fig:example} 
The image of the old open cluster, NGC 188, obtained using the NUV filter, NUVN2 is shown.}
   \end{figure} 

\item[3.] The UVIT has several filters in the FUV and NUV channels which can be used to sample the spectral energy distribution
of the observed source. This capability can be used to estimate the accurate spectral properties of sources which have energy
generation due to different physical processes or sources. We have demonstrated this capability here. The field of the old open cluster, NGC 188,
has two UV bright stars, which are relatively faint in the visual bands. These stars are thought to belong to the class of stars
known as sub-dwarfs. The UVIT measurement of flux of one of these subdwarfs in NGC 188 is shown in the figure, along with the 
previous flux measurements in the UV (GALEX and UIT), visual  and near-IR pass bands (both from ground observations). The flux
estimations from the UVIT compare well with those from UIT and GALEX. The figure demonstrates that the filter systems in the UVIT
help to widely sample the flux distribution in the UV region, with the help of its filters. Please note that the figure shows flux derived from
all the FUV filters, whereas only two are shown in the NUV.  These flux
distributions can be used to compare with the models and estimate the temperature, mass and radius of the source, which is suspected to
be a binary with a hot companion.

\begin{figure} [ht]
\begin{center}
\begin{tabular}{c} 
\includegraphics[height=7cm]{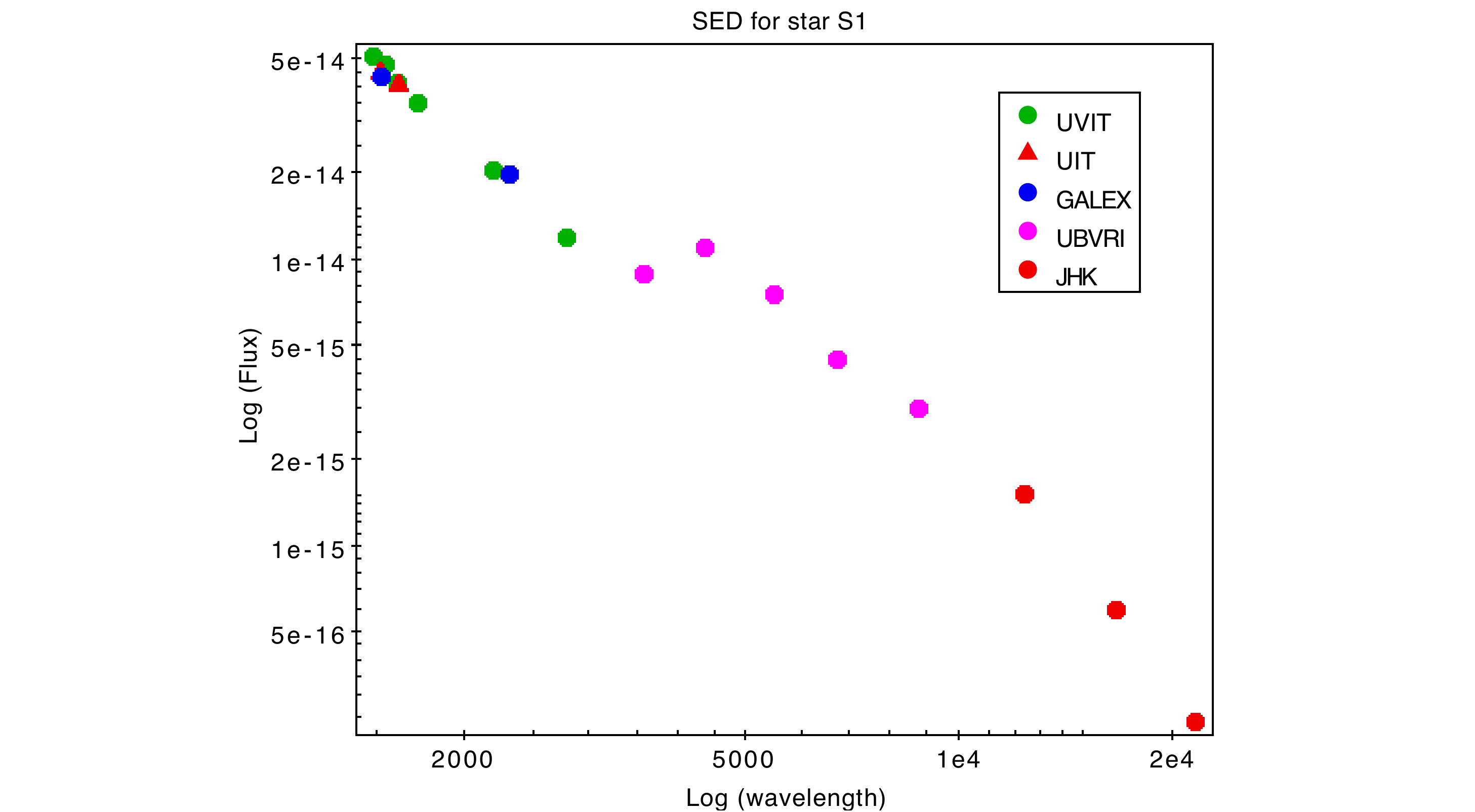}
\end{tabular}
\end{center}
\caption[example] 
   { \label{fig:example} 
The spectral energy distribution of a sub-dwarf star in the open cluster NGC 188. The flux obtained from the UVIT observations are
shown as green points.}
   \end{figure} 
\item[4.] As the detectors in the FUV and NUV channels operate in the photon counting mode, each detected photon is time tagged.
Thus, variability studies of various sources are thus possible in the FUV and NUV channels. 
Some observations, particularly AGN, were carried out to demonstrate this capability. The analysis of these are in progress.

\end{itemize}

\section{Data and Pipeline}
The UVIT data from the mission operations centre is received by the UVIT payload operation centre at IIA. The data received is
the Level1 data, in fits format, consisting of science data as well as engineering data. 
The data pipeline to create the Level2 product, where the data is ready for performing 
scientific analysis, is getting ready. In the meantime, the PV phase data is processed using a stand alone software to correct for the satellite drift and create images. These images are then used to perform various analysis to derive calibration parameters. In the next few months, the data processing is expected to become streamlined and the Level2 products are expected to be made available.

\section{Conclusions}

The In-orbit observations of the UVIT onboard the space observatory, ASTROSAT, has successfully completed the in-orbit performance
verification phase. The conclusions are summarised below:
\begin{itemize}
\item[1.] All on-board systems are performing well.
\item[2.] All filters and gratings in the three channels are tested and are found to be working fine.
\item[3.] The photometric calibration of most of the filters are completed.
\item[4.] The spectroscopic calibrations are also performed.
\item[5.] Images obtained during the PV phase demonstrate its unique capabilities, as per (if not better than) the design requirements.
\end{itemize}

\acknowledgments 
 
This publication makes use of the efforts of a large team involved with the UVIT from its conception to the delivery of the payload.
Several Indian Institutions and the Canadian Space Agency are involved in this mission and the efforts, support and funding from
all sources are acknowledged. Thanks are due to the mission team, L1 data processing team, data pipeline team, which are part of the
Indian Space Research Organisation. The efforts of the Payload operations team in IIA is also acknowledged.

{\bf References:}\\
\bibliography{report} 
\bibliographystyle{spiebib} 

Hutchings J. B., Postma J., Asquin D., Leahy D., 2007, PASP, 119, 1152\\
 Postma J., Hutchings J. B., Leahy D., 2011, PASP, 123, 833\\
 Narra S. V., Stalin C. S., Sriram S., Amit K., Tandon S. N., Determination of Quantum Efficiency of UVIT Flight Model Detectors, June 2011a (Technical Reports)\\
Sankarasubramanian K., Sreejith P., Vaishali S., Sreekumar P., Sriram S., Tandon S.N., Pradeep, June 2011, Technical Report on UVIT Filter Calibration\\
 Narra S.V., Stalin C. S., Sriram S., Pradeep R., Tandon S. N., Calibration of Engineer ing and Flight model gratings for the Ultraviolet Imaging Telescope (UVIT), June 2011b (Technical Reports)\\
 Oke, J.B. 1974, ApJS, 27, 21\\
\end{document}